\newif\ifmnras
\mnrastrue
\ifmnras
	\documentclass[fleqn,usenatbib]{mnras}
\else
	\documentclass[apj,numberedappendix]{emulateapj}
\fi

\usepackage{graphics,epsf}
\usepackage{amsmath}                
\usepackage{amsfonts}               
\usepackage{amssymb}                
\usepackage{epsfig}                 
\usepackage{graphicx}

\def \AU{~\rm{AU}}

\ifmnras
	\def \aap{A\&A}
	
	\def \apj{ApJ}
	\def \apjl{ApJ}
	\def \apjs{ApJS}

	\def \mnras{MNRAS}

\fi

\usepackage{xcolor}
\definecolor{redak}{rgb}{0.9,0.15,0.05}

\ifmnras

\title{Accounting for planet-shaped planetary nebulae}

\author	[Sabach \& Soker]{
Efrat Sabach and Noam Soker\thanks{E-mail: efrats@physics.technion.ac.il, soker@physics.technion.ac.il}
\\
Department of Physics, Technion -- Israel
Institute of Technology, Haifa 32000 Israel\\
}

	\date{Accepted XXX. Received YYY; in original form ZZZ}

	\pubyear{2017}
\fi

\begin{document}
\label{firstpage}

\ifmnras
	\pagerange{\pageref{firstpage}--\pageref{lastpage}}
	\maketitle
\else
	\title{?}

	\author{Efrat Sabach}
	\author{Noam Soker}
	\affil{Department of Physics, Technion -- Israel
	Institute of Technology, Haifa 32000, Israel;
	efrats@physics.technion.ac.il; soker@physics.technion.ac.il}
\fi

\begin{abstract}
By following the evolution of several observed exoplanetary systems we show that by lowering the mass loss rate of single solar-like stars during their two giant branches, these stars will swallow their planets at the tip of their asymptotic giant branch (AGB) phase. This will most likely lead the stars to form elliptical planetary nebulae (PNe). Under the traditional mass loss rate these stars will hardly form observable PNe. Stars with a lower mass loss rate as we propose, about 15 per cent of the traditional mass loss rate of single stars, leave the AGB with much higher luminosities than what traditional evolution produces. Hence, the assumed lower mass loss rate might also account for the presence of bright PNe in old stellar populations. We present the evolution of four exoplanetary systems that represent stellar masses in the range of $0.9-1.3 M_\odot$. The justification for this low mass loss rate is our assumption that the stellar samples that were used to derive the traditional average single-star mass loss rate were contaminated by stars that suffer binary interaction. 
\end{abstract}



\section{INTRODUCTION}
\label{sec:intro}

\subsection{Planetary Nebulae}
\label{subsec:PNe}

Observations and their interpretation strongly support the notion that most planetary nebulae (PNe) are shaped by a stellar companion that strongly interacts with the asymptotic giant branch (AGB) progenitor (e.g., limiting the list to some recent papers , \citealt{Chiotellisetal2016, DeMarcoetal2016, GarciaRojasetal2016, Harveyetal2016, Heoetal2016, Hillwigetal2016a, Hillwigetal2016b, Jones2016, Jonesetal2016, JonesBoffin2017, Chenetal2017, Macdonaldetal2017, SanchezContrerasetal2017, Miszalskietal2017, Hillwigetal2017, Sowickaetal2017}). 
PN catalogs (e.g., \citealt{Balick1987, Chuetal1987, CorradiSchwarz1995, Manchadoetal1996, SahaiTrauger1998, Sahaietal2011, Parkeretal2016}) show that most PNe are not spherical, but rather bipolar or ellipticals, suggesting that most AGB stars that formed bright enough PNe to be observed have interacted with a companion.
For more than 20 years theoretical studies have been suggesting that planets can also shape many PNe (e.g.,  \citealt{Soker1996, Nordhausetal2010, DeMarcoSoker2011}). 
  
Yet there is a problem for planets to shape PNe. 
For a planet to influence the mass loss from an AGB star, the envelope mass cannot be too large.
In addition, the planet should interact with the AGB star on the upper AGB. 
These imply an envelope mass on the upper AGB of at most several$\times 0.1 M_\odot$.
If the star mass is not too low, with a zero age main sequence (ZAMS) mass of $M_{\rm 1,ZAMS} \ga 1.4 M_\odot$,  the orbit of the planet substantially increases by the time the star losses about third of its mass, $\approx 0.5 M_\odot$.
Tidal interaction can bring it closer to the star, but one should make sure the planet is not engulfed during the red giant branch (RGB) phase  (e.g., \citealt{Nordhausetal2010}).
If the star is a low-mass star, $M_{\rm 1,ZAMS} \la 1.4 M_\odot$,  its radius on the RGB is not much smaller than that on the AGB, and if the planet is not engulfed on the RGB, it will also not be engulfed on the AGB.  

The maximum radius from which tidal forces cause a companion to spiral-in to the envelope of the giant star is called the tidal maximum capture radius. 
\cite{Madappattetal2016} conduct a thorough study of the tidal maximum capture radius of  RGB and AGB stars with initial masses of  $0.8 M_\odot < M_{\rm 1,ZAMS} \la 4 M_\odot$. 
We will not repeat their study, but rather point out the consequences of a much lower mass loss rate of RGB and AGB low mass primary stars.  
 
We will concentrate on the majority of PNe that are formed by post-AGB evolution, because one of our goals is to explain the brightest PNe in old populations. These PNe must come from post-AGB evolution.
However, we do note the possibility of post-RGB PNe, e.g., the Boomerang Nebula \citep{Sahaietal2017}. 
\cite{Hillwigetal2017} study possible post-RGB PNe, and list five candidates. They note that though the fraction of post-RGB PNe is estimated to be very low among all PNe, it might possibly be higher than present estimates.  

We are most interested in planets shaping PNe with an initial progenitor mass in the range of $M_{\rm 1,ZAMS} \simeq 0.9-1.3 M_\odot$.
For these stars the companion, whether a star, a brown dwarf, or a planet, has also the role of increasing the mass loss rate and expediting the evolution on the final AGB and the post-AGB phases.
This enhanced mass loss rate is required for the PN to be bright enough to be observed \citep{DeMarcoMoe2005, SokerSubag2005, MoeDeMarco2006}.  
But, in the commonly used mass loss rate these stars reach an RGB radius that is not much below, and even larger, than their maximum radius on the AGB.
If, on the other hand, the low-mass star does not interact with a companion on the RGB, it will most likely not interact on the AGB. 
 
This scenario will vary if the mass loss rate on the RGB and AGB is much lower than what is usually used in stellar evolution codes, since the star then reaches a much larger radius on the AGB.
Our goal is to explore the implications of a much lower mass loss rate on the giant branches of low mass stars on such systems.

\subsection{Mass loss}
\label{subsec:massloss}

A stellar companion at a close orbit, or inside the envelope, deposits angular momentum and energy to the envelope of a giant star and by that increases the mass loss rate. 
Low mass RGB and AGB stars are influenced not only be stellar, but also by sub-stellar (i.e., brown dwarfs and planets) companions that can deposit a substantial amount of angular momentum to the envelope when they are swallowed by the expanding giant star 
(e.g., \citealt{Soker1996, Carlbergetal2009, VillaverLivio2009, MustillVillaver2012,
NordhausSpiegel2013, GarciaSeguraetal2014, Staffetal2016, AguileraGomezetal2016}).
In those cases deposition of angular momentum is more significant than
deposition of energy, e.g., for the operation of a dynamo in the envelope
of the giant star (e.g., \citealt{NordhausBlackman2006}).
\cite{Soker2004IAUS} summarized the processes by which a planet can influence the mass loss rate and geometry from RGB and AGB stars. The main process is enhanced dust formation on the surface, that in turn facilitates the usage of the stellar luminosity to remove mass. 

Since most stars with ZAMS mass of $M_{\rm ZAMS} \ga 1 M_\odot$ have close companions, stellar or sub-stellar (e.g., \citealt{Bowleretal2010}), most stars suffer strong binary interaction before they turn to a white dwarf (WD).
In many cases, in particular when the companion is of low mass and is swallowed by the star, the companion will not survive the evolution, and will leave a single WD.
We therefore conclude that most giant stars for which the mass loss rate has been determined, whether directly or statistically from the relation between the ZAMS mass and the WD mass, suffered strong binary interaction. 
In other words, the mass loss rate formulae of RGB and AGB stars include in them a substantial component of binary interaction. 
We here take the view that the mass loss rate on the RGB and AGB of stars that did not suffer any binary interaction is much lower than what traditional fitting formulae suggest.
  
 We are aware that this is a speculative view, but nonetheless, because of the far reaching implications of this assumption, we explore one of its consequences.
We further discuss the reduced mass loss rate in section \ref{sec:reducedmassloss}. 
  
\cite{Villaveretal2014} study the influence of the mass loss rate on the fate of close planets. We differ from them in three significant aspects. Firstly, they do not follow the system to the AGB phase. Secondly, they lower the mass loss rate to 40 per cents of the regular mass loss rate assumed ($\eta=0.2$).
We, based on our claim that the determination of the mass loss rate on the RGB and AGB is contaminated with many stars that suffer binary interaction with a stellar or a substellar object,
explore the possibility that the mass loss rate might be as low as around 15 percent the commonly used value.
Thirdly, we concentrate on lower mass stars $M_{\rm ZAMS} \simeq 0.9-1.3 M_\odot$, while \cite{Villaveretal2014} study the range of $M_{\rm ZAMS} = 1.5-2 M_\odot$. 
These differences are significant as we find qualitatively very different results for the two mass loss rates (commonly used mass loss rate and lowered mass loss rate). 
  
\subsection{The fate of known exoplanets}
\label{subsec:exoplanets}

The interaction of planets with stars evolving off the main sequence has been the focus of many studies for over two decades (e.g., \citealt{Soker1996, SiessLivio1999a, SiessLivio1999b, VillaverLivio2007, VillaverLivio2009, Nordhausetal2010, DeMarcoSoker2011, Mustilletal2014, Villaveretal2014, Meynetetal2017}, to list a small sample of all relevant papers). 
\cite{NordhausSpiegel2013} are different than most studies in that they target about 300 known exoplanets. Their figure 4 very nicely summarizes their results. They take the commonly used mass loss rate. Eventually, a follow-up study of our present preliminary study will have to repeat the calculations of \cite{NordhausSpiegel2013} but using a lower mass loss rate. We do note that if there are two planets and the inner one is swallowed by the giant, then the mass loss rate increases, hence increasing the survivability of the further out planets. 

We will not repeat the calculations that have been performed in those papers, but use some of the results. 
In particular, because of several uncertainties in the tidal capturing process, and our claim here for uncertainties in the mass loss rate, there is no need in the present preliminary suggestion of the low-mass loss rate to perform an accurate integration of the equations of motion during the tidal capture process.
There are uncertainties in the tidal efficiency itself and there is much debate on the exact mechanism and resulting capture radius.
\cite{Madappattetal2016} integrate the tidal equations of \cite{Zahn1977, Zahn1989} and find that low mass AGB stars with an initial mass in the range of $0.8-4M_\odot$ capture companions that are between 1 and 4 times the maximum giant radius.
\cite{VillaverLivio2009} study the planet orbit along the RGB phase of stars in the mass range of $1-5M_\odot$ with a \cite{Reimers1975} mass loss efficiency parameter  $\eta=0.6$, 
and find that under their assumptions a $1 M_{\rm J}$ planet companion, where $M_{\rm J}$ is the mass of Jupiter, will be engulfed by a $1 M_\odot$ star at a distance of $a<3 \AU$.
\cite{MustillVillaver2012} follow the thermal pulses along the AGB phase of low mass stars and their effect on a planet companion. They find that for a $1M_\odot$ primary tidal forces, calculated with the \cite{Zahn1977} formalism, are strong enough to pull giant planets at an orbital separation of $3 \AU$.

Here, we crudely estimate from the results of \cite{Soker1996} and of \cite{Villaveretal2014} that a planet will be tidally captured if the planet orbital separation when the star is on the main sequence is
\begin{equation}
a_i < a_{\rm cap} \simeq 3.1 R_{\rm \ast, max} 
\left( \frac{m_p}{0.002M_1} \right)^{1/8} , 
\label{eq:acapture}
\end{equation}
where $R_{\rm \ast, max}$ is the maximum radius of the RGB or AGB star (not including pulsations), $M_1$ is the mass of the giant star, and $m_p$ is the mass of the planet or brown dwarf. 
The inequality (\ref{eq:acapture}) is basically equation (6) of \cite{Soker1996}, that is calibrated for the case of a planet mass equal to $2M_J$, and interacting with a $\approx 1M_\odot$ RGB or AGB star.
For eccentric orbits the value of $a_{\rm cap}$ becomes even larger. The value of the capture orbital separation $a_{\rm cap}$ weakly depends also on other parameters (see the relevant references), but these are not significant for our study. 

\section{REDUCED MASS LOSS RATE}
\label{sec:reducedmassloss}
In this study we will not discuss the mass loss mechanisms that are the topics of many reviews 
(e.g., \citealt{Lafon&Berruyer1991, SchroderCuntz2005}).
We here only apply our assumption that the mass loss rate of giant stars that did not suffer any interaction with a massive planet, a brown dwarf, or a stellar companion, is very low.
For that we will consider the Reimers empirical mass loss rate for red giant stars \citep{Reimers1975}, that can be written as 
\begin{equation}
\dot{M}=\eta\times4\times10^{-13} LM^{-1}R,
\label{eq:Reimers}
\end{equation}
where the stellar luminosity $L$, mass $M$, and radius $R$ are in solar units, 
and $\eta$ is the mass loss rate efficiency parameter that is determined from observations.

\cite{McDonaldZijlstra2015} conduct a detailed and thorough study on the value of
the mass loss rate on the RGB in globular clusters.
They use the horizontal branch (HB) morphology to deduce the value of $\eta$, and
find a median value of $\eta=0.477$.
Below we will use $\eta=0.5$ to follow the evolution of stars under the commonly assumed mass loss rate (e.g., \citealt{Guoetal2017}).
We study the effects of a reduced mass loss rate efficiency parameter 
for stars that did not suffer any interaction with a companion
and present the results for a representative value of $\eta=0.07$.
We do not see a contradiction between our assumption and the results of
\cite{McDonaldZijlstra2015} for the following reasons.
(1) We agree that the typical value of $\eta$ for solar type stars is around $\approx 0.5$.
Yet, we claim that this typical value includes many RGB and AGB stars that suffer
interaction with a companion, being stellar or sub-stellar. 
(2) Their mass loss formula with $\eta \simeq 0.5$ cannot cover the entire stellar population on the HB. In particular it cannot cover bluer HB stars.
Indeed, already \cite{DCruz1996} noted that a range of values of $\eta$ is required
to produce the population of stars on the HB
(unless helium abundance forms it). 
(3) There are observations that the study of \cite{McDonaldZijlstra2015} cannot account for.
In particular some bright AGB stars, up to $5000 L_\odot$.
It is exactly those bright AGB and post-AGB stars of low mass progenitors that our proposed low mass loss rate intends to explain, as well as the shaping of PNe by planets. 

We justify our usage of a low mass loss rate for giant stars that suffer no binary interaction by restating our basic claim as follows.
The sample of giant, or post-giant, stars that have been used to deduce the semi-empirical mass loss rate formulae in different studies (in open clusters, in globular clusters, in the field) are substantially contaminated with stars that did suffer interaction with stellar and sub-stellar objects.
These companions enhance the mass loss rate by an appreciable factor. In most cases of claimed single-star evolution, the companion that enhanced the mass loss rate did not survive the binary evolution, e.g., it was tidally destructed or it merged with the core of the giant star.    
 
We start by studying stellar models of low mass stars with ZAMS mass $M_{\rm 1,ZMAS}\la 1.3M_\odot$ , and for each compare
the evolution with the commonly used mass loss to the evolution with reduced mass loss.
For the commonly used mass loss rate we take the typical value of the Reimers parameter for solar type stars $\eta \approx 0.5$, as mentioned above.
For the reduced mass loss rate we present the evolution with a representative case of a much lower mass loss rate of $\eta=0.07$, both for the RGB and the AGB.
In section \ref{sec:evolution} we explain the reason for presenting the case of $\eta=0.07$ among the several different values that we have studied (see full results in the Appendix).

The notion of a much lower mass loss-rate is not new.
\cite{Miglioetal2012}, for example, find that for the old metal-rich cluster
NGC 6791 the red giant mass loss rate should be lower than the typically taken, where
the efficiency parameter might be as low as $\eta=0.1$ and up to $0.3$.
For NGC 6819 they find that the RGB efficiency parameter could be very low, yet as it is a young cluster the constraints on $\eta$ are less compelling.

We differ in that we conduct a systematic comparison, and attribute the low
mass loss rate to stars that suffered no interaction with a companion.  

\section{THE SAMPLE OF EXOPLANETS}
\label{sec:exoplanets}

To study the effect of a low mass loss rate of low mass stars, $M_{\rm 1,ZMAS}\la 1.3M_\odot$,
on the fate of planets we perform stellar evolution simulations of a sample of observed exoplanetary systems. 
We take the systems from The Extrasolar Planets Encyclopaedia, 
\url{http://exoplanet.eu/catalog/} \citep{Schneider2011}, according to the following criteria. 
The star mass is in a mass range of $M_{\rm 1,ZMAS}:0.9- 1.3M_\odot$ with metallicity of about solar, $Z=0.02$, the planet mass is in a mass range of $m_p\simeq 1-10M_J$,
low eccentricity, and a semi-major axis of $a_i\simeq 2-5 R_\odot$.

In Table \ref{tab:planets} we list the four exoplanetary systems that we present here. 
We preset these 4 systems because the mass of their stars span the mass range of interest, $M_{\rm 1,ZMAS}:0.9- 1.3M_\odot$ and the orbital separations of their planets demonstrate the effects we study. We point out that these four systems are not unique, and that there are other exoplanetary systems that are compatible with our criteria (e.g., HD 72659 b, HD 108874 c,  HD 222155 b, and more), but they will not shed new light on the phenomena we study.  
\begin{table*}
    \centering          
    \begin{tabular}{lcccccc }
	\multicolumn{7}{ c }{} \\	    
    \hline              
    \hline              
    & Planet name  & $M_i$       & $Z$ &  $a_i$ &   $e$ &  $m_p$   
\\  &             & $[M_\odot]$ &     & $[AU]$ &       & $[M_J]$
     \\[1ex]
    \hline              
  (a) & HD 290327 A b  & 0.9       & 0.015 & 3.43  & 0.08   & 2.5
\\(b) & 47 Uma b       & 1.03      & 0.02  & 2.10  & 0.03   & 2.5
\\(c) & HD 159868 b    & 1.09      & 0.02  & 2.25  & 0.01   & 2.1
\\(d) & HD 13908  c    & 1.29      & 0.02  & 2.03  & 0.12   & 5.1 \\
    \hline     
    \end{tabular}
\footnotesize
\flushleft
\caption{Some properties of the four exoplanetary systems we study.
The parameters listed are the planet name, the initial mass of the primary star $M_i$, metallicity $Z$, the orbital semi-major axis $a_i$, eccentricity, and the mass of the planetary companion $m_p$ in units of Jupiter mass $M_j$. 
Sources for the properties of these systems are as follows. 
HD 290327 b \citep{Naefetal2010}, 47 Uma b \citep{Fuhrmannetal1997}, HD 159868 b 
\citep{Wittenmyeretal2012}, and HD 13908 c \citep{Moutouetal2014}.
}
\label{tab:planets}
\end{table*}

\section{EVOLUTION}
\label{sec:evolution}
We use the stellar evolution code \texttt{MESA} (Modules for Experiments in Stellar Astrophysics)  version 9575 (\citealt{Paxtonetal2011, Paxtonetal2013, Paxtonetal2015}).
For each of the stars in our sample (see section \ref{sec:exoplanets}) we calculate the stellar evolution from ZAMS until the formation of a WD and study the differences between two evolutionary paths, one with regular and one with low mass loss rate as described in section \ref{eq:Reimers}.
The Reimers mass loss prescription is taken for the RGB and the prescription of \cite{Bloecker1995} is taken for the AGB.
We are interested in determining the fate of the planet, namely whether the star will swallow the planet on the RGB or on the AGB, or not at all. 

We studied the evolution with six different values of the mass loss rate parameter, 
$\eta=0.5$, $0.35$, $0.25$, $0.15$, $0.07$, and $0.05$. We found that the processes we study here, of planets interacting with AGB stars and bright post-AGB stars, are obtained in most cases for $\eta=0.15$. For example, already for $\eta =0.15$ the maximum radius on the AGB is significantly larger than that on the RGB.
To clearly demonstrate these processes, however, we chose to present here the results for  
a representative value of $\eta=0.07$. We somewhat arbitrarily chose this value as it falls in the relevant range, $0.05\lesssim \eta \lesssim 0.15$, where the effects of a reduced mass loss rate are more pronounced than for higher values of $ \eta > 0.1$. 
We compare the $\eta=0.07$ evolution with the commonly used $\eta=0.5$ value. In the Appendix we compare the evolution of the six different values of $\eta$ and present the full results.

\begin{figure*}
\centering
{(a)}
\includegraphics
[trim= 3cm 6cm 4.6cm 5cm,clip=true,width=0.45\textwidth]{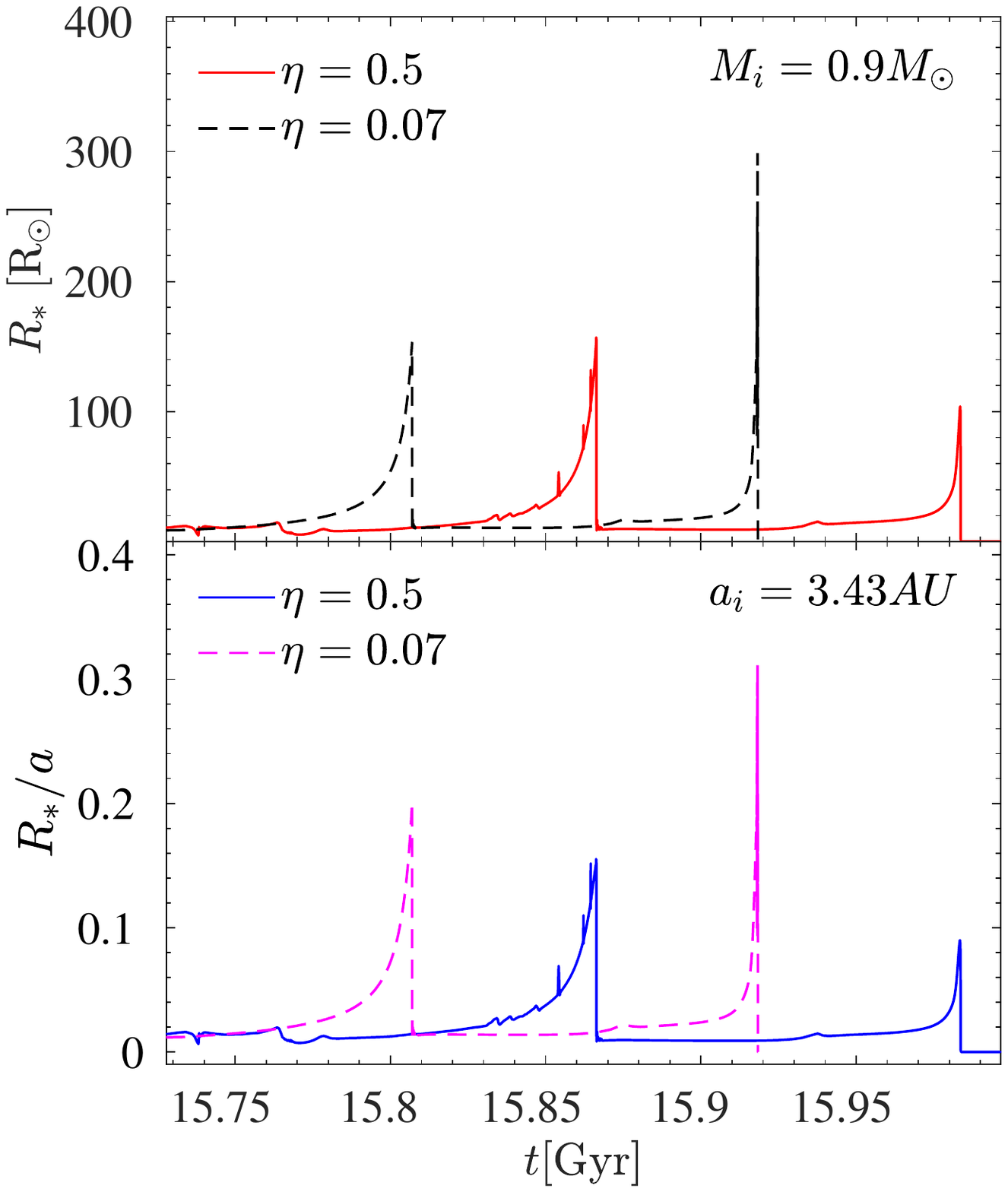} 
\hspace*{0.5cm}
\vspace*{0.5cm}
{(b)}
\includegraphics
[trim= 3cm 6cm 4.6cm 5cm,clip=true,width=0.45\textwidth]{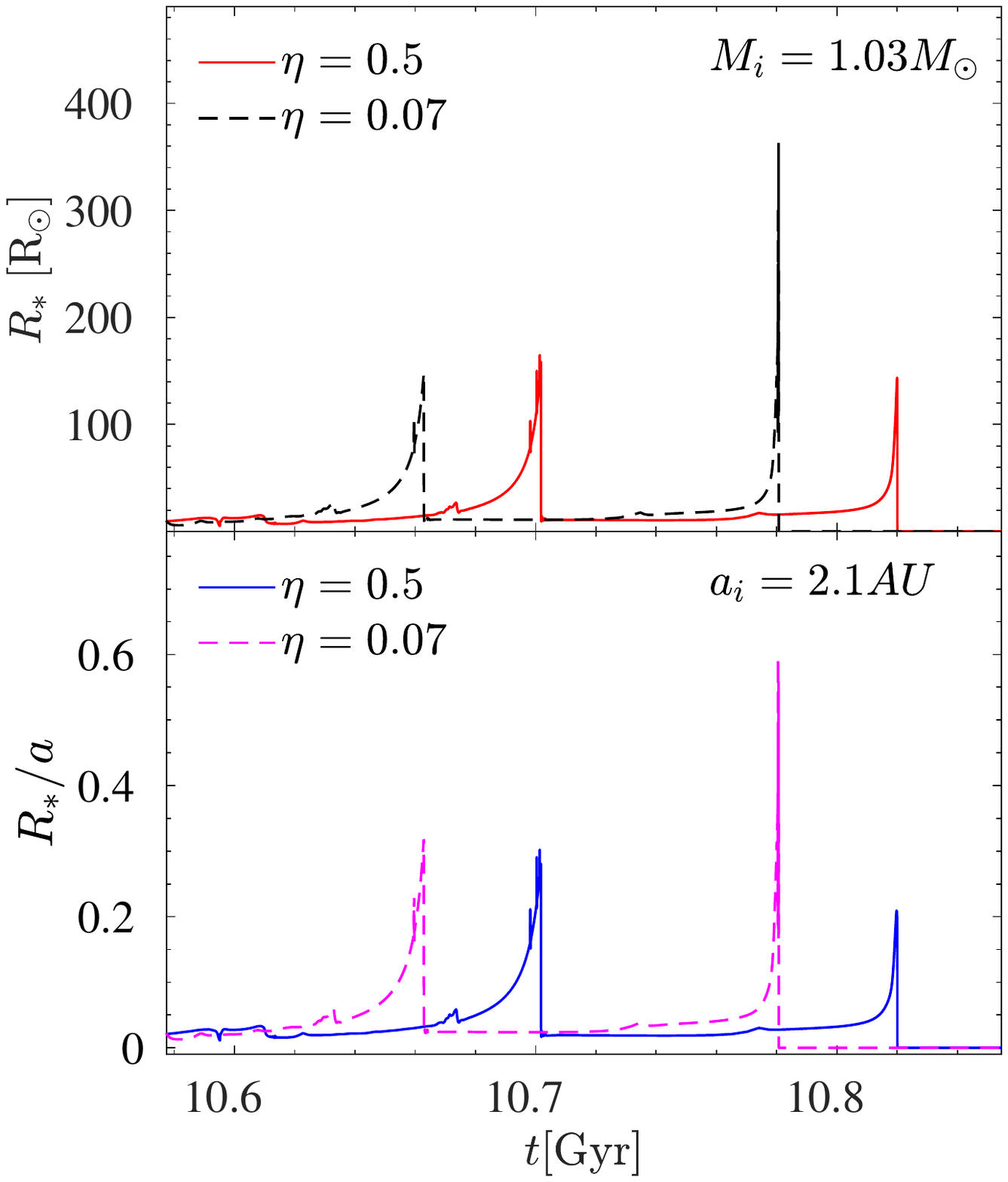} 
{(c)}
\includegraphics
[trim= 3cm 6cm 4.6cm 5cm,clip=true,width=0.45\textwidth]{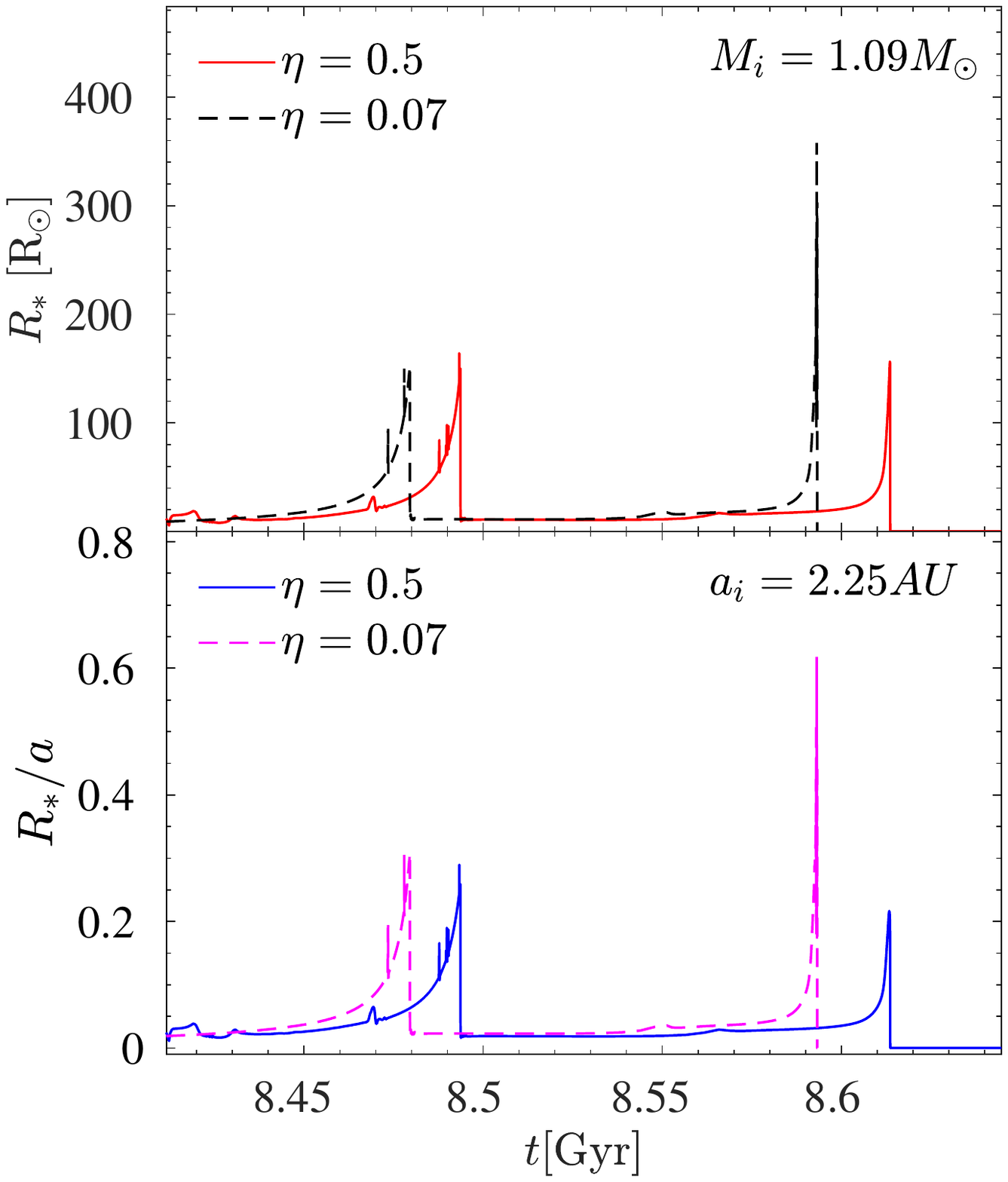} 
\hspace*{0.5cm}
{(d)}
\includegraphics
[trim= 3cm 6cm 4.6cm 5cm,clip=true,width=0.45\textwidth]{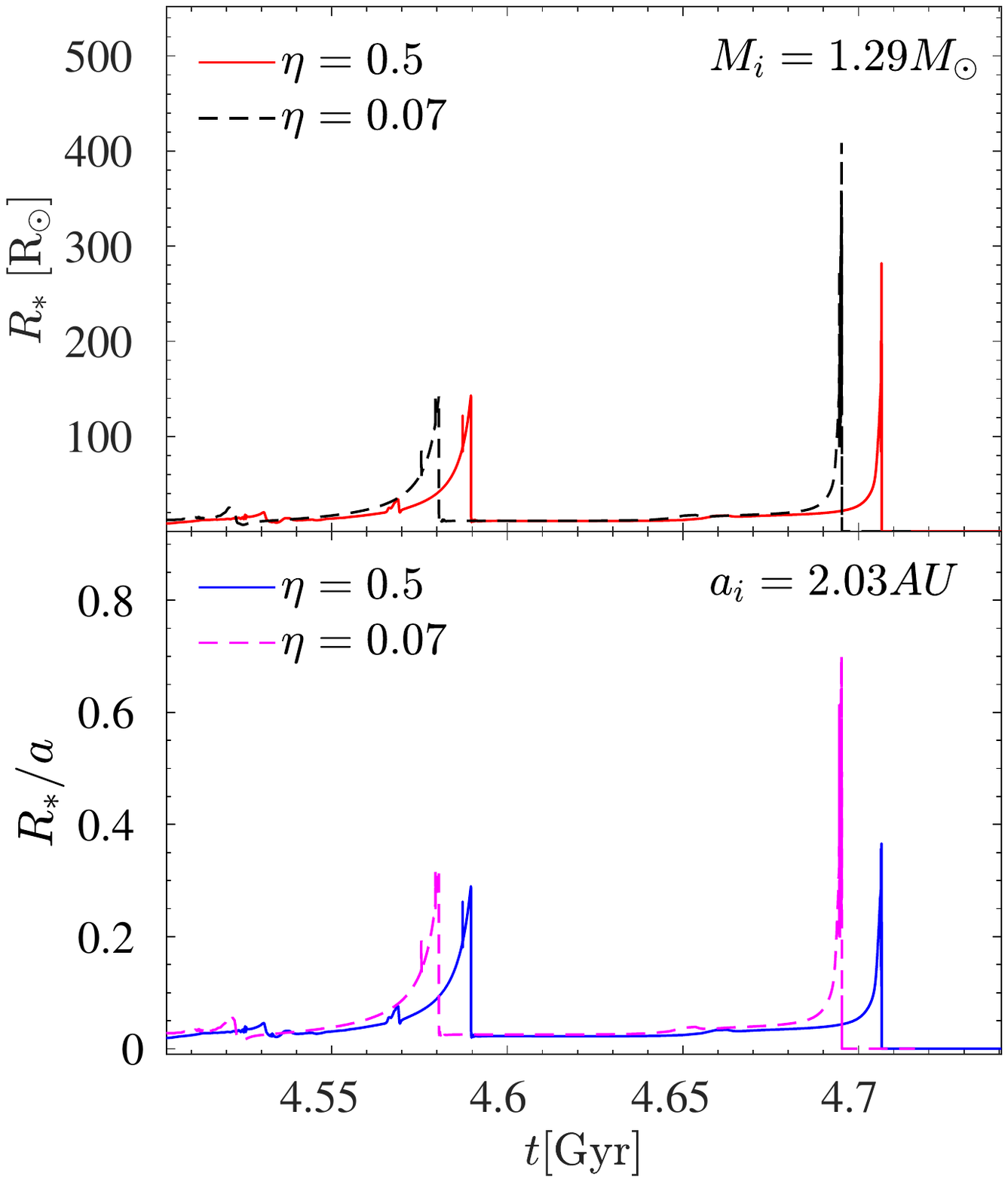} 
\vspace*{0.5cm}
\caption
{Evolution of four stellar systems that are listed in Table \ref{tab:planets}. We use \texttt{MESA} to follow the stellar evolution until the formation of a WD in each case.
For each of the four exoplanetary systems we indicate in the panels the initial stellar mass $M_i$ and the initial semi-major axis of the planetary orbit $a_i$.  
In the upper panel of each case we present the evolution of the stellar radius with time for the commonly used mass loss rate of $\eta=0.5$ (solid-red line), and for a representative reduced mass loss rate of $\eta=0.07$ (dashed-black line). The time is given from the main sequence.
In the lower panel of each case we show the ratio of the stellar radius to the orbital separation for the two mass loss rates. In calculating the orbital separation we include the stellar mass loss process, but not tidal interaction. 
}
\label{fig:planets}
\end{figure*}

In Fig. \ref{fig:planets} we present the stellar radius during the post-main sequence evolution, $R_\ast$, and the ratio of the stellar radius to the orbital separation (semi-major axis) of the planet, for the four exoplanetary systems that we list in Table \ref{tab:planets}.  The solid lines in all panels are for the evolution with the 
the commonly used mass loss rate, $\eta=0.5$ in equation (\ref{eq:Reimers}), while the dashed lines depict the evolution with a reduced mass loss rate with a representative efficiency parameter of $\eta=0.07$.
In calculating the evolution of the orbital separation $a$ we consider only the mass loss process, and do not include tidal forces (as we explained in section \ref{subsec:exoplanets}). 

Let us apply the approximate expression for the tidal maximum capture radius
(eq. \ref{eq:acapture}) to our results. We can express the capture condition as $R_{\rm \ast, max}/a_i \ga 0.32$, with large uncertainties.  
Since the orbital separation increases only slightly before the low mass star reaches the upper AGB, this condition can be written also for the evolving orbital separation as $R_{\rm \ast, max}/a \ga 0.3$, again, with large uncertainties.  

It is evident from our results that for the case of evolution with the commonly used mass loss rate, solid lines in Fig. \ref{fig:planets}, our sample stars will not (for $M_i=0.9 M_\odot$) or only marginally, if at all, engulf their planet companions during the RGB phase. Moreover, if the planet is not engulfed during the RGB phase, it will not be engulfed during the AGB phase, or barely so for the $M_i=1.29 M_\odot$ case. 
In other words, for the commonly used mass loss rate the orbital separation for which planets can be engulfed during the AGB phase of low mass stars is tiny or doesn't exist.  
If the planet is engulfed on the RGB it will increase the mass loss rate, and the star will not reach the AGB at all, or only the lower part of the AGB. No PN will be formed. 
If the planet is not engulfed at all, then it is most likely that the star will form a spherical and very faint PN (e.g., \citealt{SokerSubag2005, DeMarcoMoe2005, MoeDeMarco2006, DeMarco2009}).

The fate of the systems in the cases with the low mass loss rate can be much different. In these cases the maximum stellar radius on the AGB becomes much larger than that on the RGB.
There is a relatively significant range of initial orbital separations for which the capture condition $R_{\rm \ast, max}/a \ga 0.3$ does not hold on the RGB (or only marginally so), while it is fulfilled on the AGB. This implies that the planet will be engulfed while the star is on the upper AGB. The engulfed planet will enhance the mass loss rate and make the nebula denser and elliptical. Overall, this evolution can lead to an observed PN. 

In Table \ref{tab:results} we present some properties of the exoplanetary systems that we evolved. We note that the lower mass loss rate implies also that the luminosity of the central star is larger than in the commonly used mass loss rate (column 9). It is therefore possible that some of the PNe will be [OIII] bright, despite that they result from stars of only $M_i \simeq 1-1.2 M_\odot$. This might contribute to the explanation of why some bright PNe are observed even in old stellar populations (e.g., \citealt{Ciardullo2010}).
\begin{table*}
    \centering          
    \begin{tabular}{lccccccc}
	\multicolumn{8}{ c }{} \\	    
    \hline              
    \hline              
    Planet name & $M_i$    & $a_i$ & $m_p$  & $\eta$  & $\left(^{R_*}/_{a}\right)_{\rm max}$ &  $f_{\rm AR}$ & $L_{\rm pAGB}$
\\             &$[M_\odot]$&$[AU]$ & $[M_J]$&         &  RGB ~~ AGB     					 &               & $[L_\odot]$
     \\[1ex]
    \hline              
  HD 290327 A b &  0.9    & 3.43   & 2.5    &	0.5   &  0.15 ~~ 0.09						 & 0.65			 & $1.47\times10^3$
\\HD 290327 A b &  0.9    & 3.43   & 2.5    &	0.07  &  0.20 ~~ 0.31						 & 1.22			 & $4.23\times10^3$ 
\\47 Uma b      &  1.03   & 2.10   & 2.5    &   0.5   &  0.30 ~~ 0.21						 & 0.85			 & $1.97\times10^3$
\\47 Uma b      &  1.03   & 2.10   & 2.5    &   0.07  &  0.32 ~~ 0.59						 & 1.51			 & $4.62\times10^3$
\\HD 159868 b   &  1.09   & 2.25   & 2.1    &   0.5	  &  0.29 ~~ 0.22						 & 0.96			 & $2.18\times10^3$
\\HD 159868 b   &  1.09   & 2.25   & 2.1    &   0.07  &  0.30 ~~ 0.62						 & 1.53			 & $5.24\times10^3$
\\HD 13908  c   &  1.29   & 2.03   & 5.1    &	0.5	  &  0.30 ~~ 0.37						 & 1.20			 & $3.30\times10^3$
\\HD 13908  c   &  1.29   & 2.03   & 5.1    &	0.07  &  0.32 ~~ 0.69						 & 1.81			 & $5.69\times10^3$
\\ \hline
    \end{tabular}
\footnotesize
\flushleft
\caption{The results for the four exoplanetary systems we study. For each system the upper line is for the commonly used mass loss rate and the lower line is for a reduced mass loss rate. From left to right: the planet name, initial mass of the primary star $M_i$, initial semi-major axis $a_i$, mass of planetary companion $m_p$ (observed parameters), the mass loss rate parameter $\eta$ (our input parameter), and then our results, the  maximum ratio of the stellar radius to the orbital separation on the RGB and on the AGB, the calculated $f_{\rm AR}$ ratio (eq. \ref{eq:FAR}) that depicts the relative importance of tidal forces on the AGB to those on the RGB, and the post-AGB luminosity of the stellar remnant. 
}
\label{tab:results}
\end{table*}

To further emphasize the difference between the commonly used and the low mass loss rates, we examine the operation of the tidal forces. 
The rate of spiraling-in due to tidal forces depends on several properties of the star. But for a given orbital separation it mainly depends on the stellar radius to the power of 8 \citep{Zahn1977, VerbuntPhinney1995}. The efficiency of the tidal forces to bring the planet into the envelope, therefore, is about proportional to the integral of $\int R^8_* dt$.
We define the $f_{\rm AR}$ ratio as the ratio of this integral on the AGB to that on the RGB to the power of $1/8$, 
\begin{equation}
f_{\rm AR} \equiv \left[ \frac{\int_{T_{\rm AGB}} R^8(t)dt}{\int_{T_{\rm RGB}} R^8(t)dt} \right]^{1/8},
\label{eq:FAR}
\end{equation}
where $T_{\rm AGB}$ is the total lifetime on the AGB and $T_{\rm RGB}$ is the total lifetime on the RGB.
If $f_{\rm AR}> 1$, then there is a range of initial orbital separations for which a planet can survive tidal capture during the RGB phase of its parent star, and be engulfed during the stellar AGB phase. The system will form an elliptical PN in that case. 
We list the values of the $f_{\rm AR}$ ratio for the different cases in Table \ref{tab:results} (column 8).
The value of the $f_{\rm AR}$ ratio for the low mass loss rate case for each system is much larger than that of the regular mass loss rate. 

\begin{figure*}
\centering
{(a)}
\includegraphics
[trim= 3cm 6cm 4.6cm 5cm,clip=true,width=0.45\textwidth]{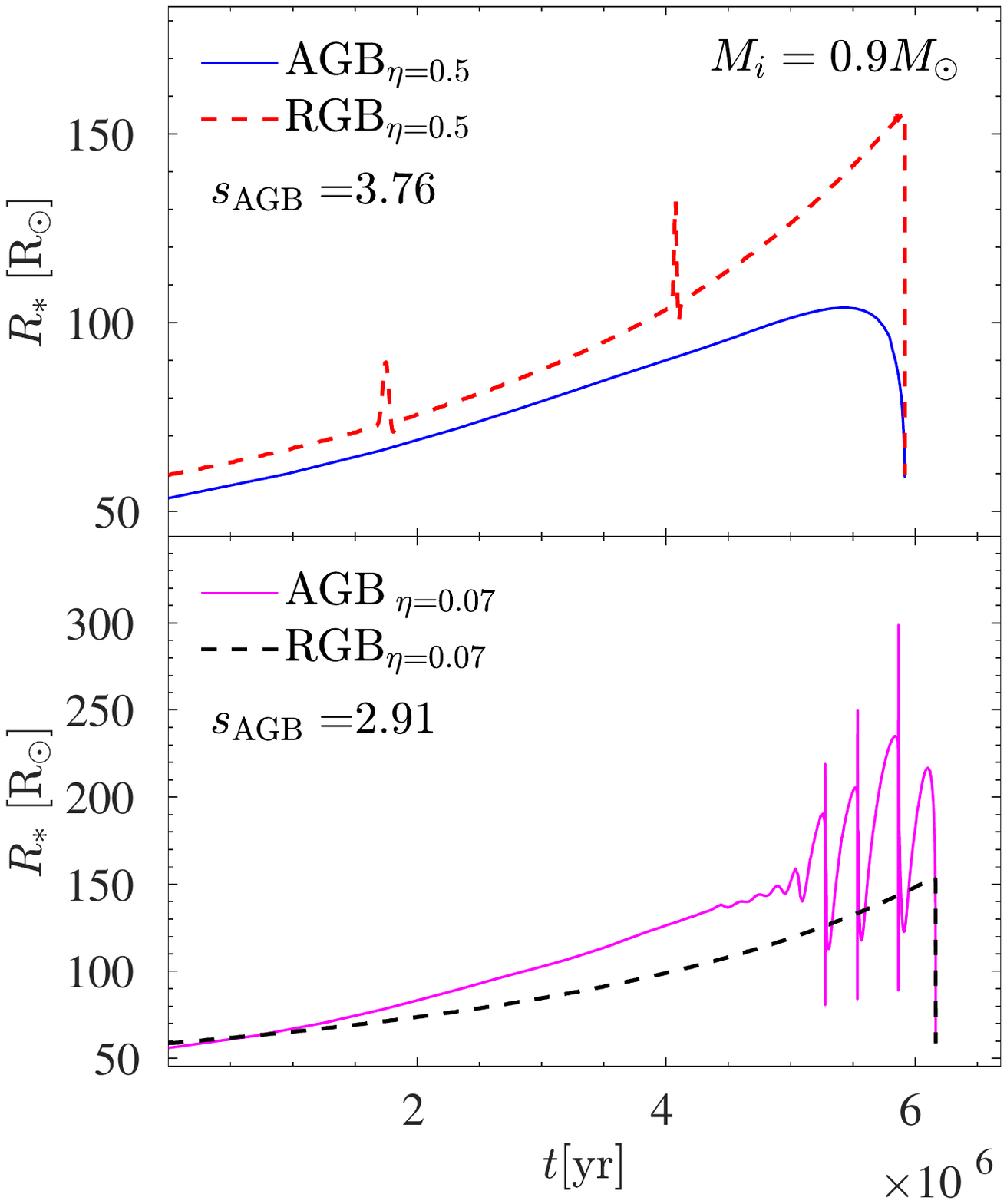} 
\hspace*{0.5cm}
\vspace*{0.5cm}
{(b)}\includegraphics
[trim= 3cm 6cm 4.6cm 5cm,clip=true,width=0.45\textwidth]{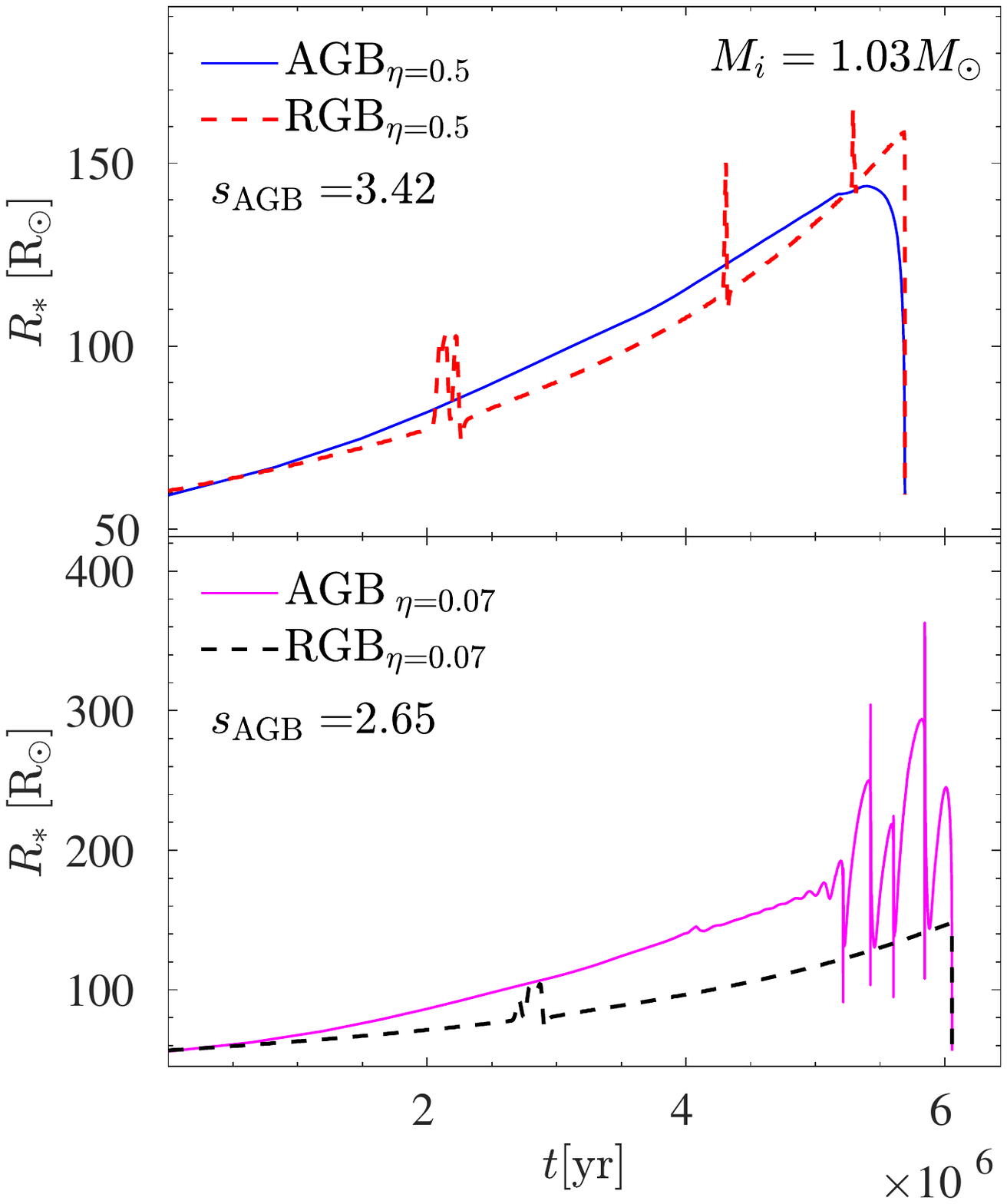} 
{(c)}
\includegraphics
[trim= 3cm 6cm 4.6cm 5cm,clip=true,width=0.45\textwidth]{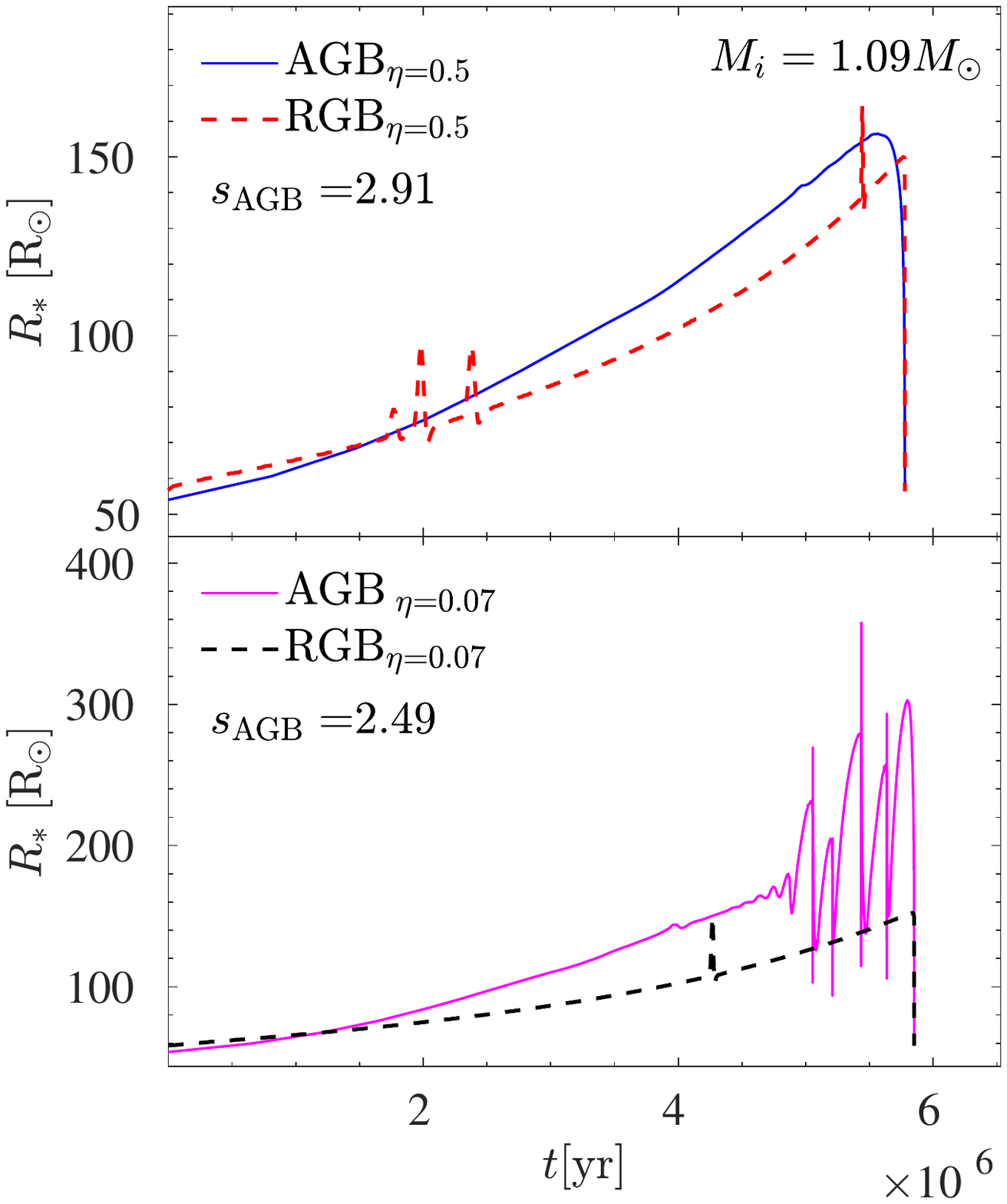} 
\hspace*{0.5cm}
{(d)}
\includegraphics
[trim= 3cm 6cm 4.6cm 5cm,clip=true,width=0.45\textwidth]{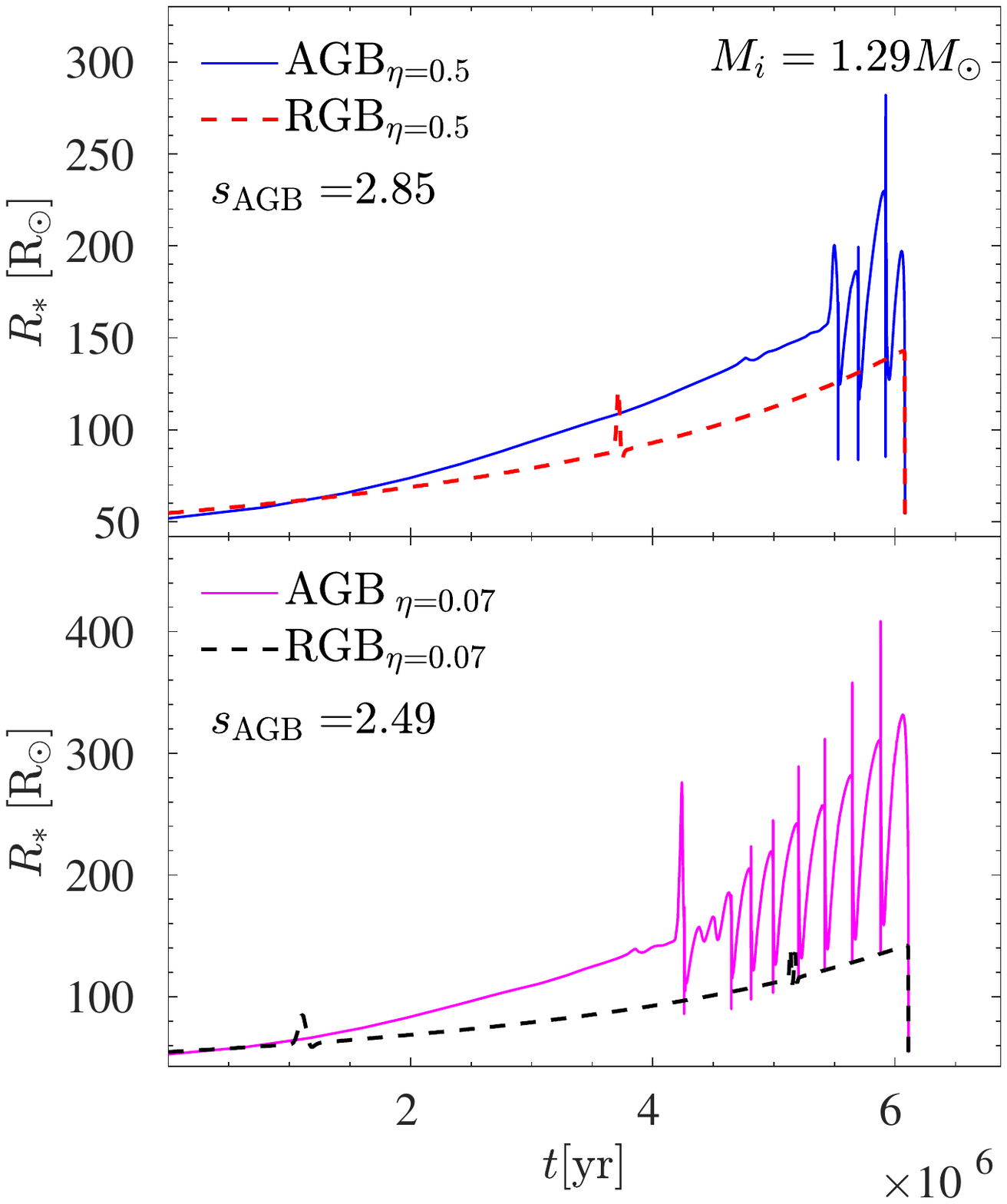} 
\caption{Comparing the evolution of the stellar radius on the RGB with that on the AGB for the four stellar models studied here, as indicated in each case for the systems in Table \ref{tab:planets}. In each case the upper panel is for the commonly used mass loss rate with $\eta=0.5$, and the lower panel is for a representative lower mass loss rate of $\eta=0.07$.
In all panels the dashed line is for the radius on the RGB and the solid line is for that on the AGB. To facilitate comparison, the AGB evolution time was stretched by a factor of $s_{\rm AGB}$, as indicated in each panel. Namely, the true AGB evolution time is that given on the axis divided by $s_{\rm AGB}$. 
The lower mass loss rate leads to more helium shell flashes and larger radii on the upper AGB. 
}
\label{fig:giants}
\end{figure*}

In Fig. \ref{fig:giants} we compare the evolution of the stellar radii on the RGB and AGB phases of each star. To facilitate such a comparison we stretch the time scale of the AGB phase by a factor of $s_{\rm AGB}$, as given in each panel. 
In the upper panel of each star we present the evolution with the commonly used mass loss rate, $\eta=0.5$, and in the lower panel we present the evolution with a reduced mass loss rate with a representative efficiency parameter of $\eta=0.07$.
For the tidal interaction in the cases of a low mass loss rate, the larger radii on the AGB have a larger effect than the longer duration of the RGB.
This ,again, shows that there is a range of orbital separations for planets to survive the RGB phase but to be engulfed during the AGB phase.   

The most significant conclusion from our results, as presented in Fig. \ref{fig:planets}, Fig. \ref{fig:giants} and Table \ref{tab:results}, 
is that for the reduced mass loss rate there is a relatively large orbital separation range for which a star can capture a planet during its AGB phase.
We showed this for four observed exoplanetary systems.
A planet of Jupiter mass, or any more massive companion, that is captured into the very extended and low mass envelope ($M_{\rm env} \la 0.5 M_\odot$) of an upper AGB low mass star will enhance the mass loss rate and will lead to the formation of an elliptical PN.
With the commonly used mass loss rate a spherical and very faint PN is expected, or no PN at all.

\section{DISCUSSION AND SUMMARY}
\label{sec:discussion&summary}
In this paper we assume that the mass loss rate of low mass giant stars, $M_{\rm 1,ZMAS}:0.9- 1.3M_\odot$, that suffer no interaction with a companion, stellar or sub-stellar, is very low.
In other words, we claim that the samples of stars that have been used in different studies over the years to derive the mass loss rate on the giant branches were contaminated with stars that suffered binary interaction. Such binary interaction substantially enhances the mass loss rate.
To study the effects of a reduced mass loss rate, we changed the mass loss parameter in equation (\ref{eq:Reimers}) and studied the evolution with six different values, 
$\eta=0.5$, $0.35$, $0.25$, $0.15$, $0.07$, and $0.05$ (see Appendix).
We found that the processes we study here, of planets interacting with AGB stars and bright 
post-AGB stars, are obtained in most cases for $0.05\lesssim \eta \lesssim 0.15$. 
To clearly demonstrate these processes we presented the results for a representative value 
of $\eta=0.07$.

We examined some consequences of our assumed low mass loss rate. We simulated the evolution of the parent stars of four exoplanetary systems (listed in Table \ref{tab:planets}).  
We followed the evolution of the stellar radii and the planetary orbital separations under the assumption of both the commonly used and the reduced mass loss rates. The results are presented in Fig. \ref{fig:planets} and are summarized in Table \ref{tab:results}
(for the entire study of the effects of different mass loss rate parameters on low mass stars see Appendix).

Our somewhat speculative assumption has several implications, including possible solutions to some riddles. 

\emph{The fraction of PNe that are shaped by planets.}
Previous studies argued that some fraction of PNe were shaped by brown dwarf or planet companions (e.g., \citealt{Soker1996} and \citealt{DeMarcoSoker2011}). The rest were shaped by stellar companions.
Under the assumed low mass loss rate, AGB stars reach much larger radii, in particular relative to their maximum radius on the RGB.
This leaves a relatively large range of initial planetary orbital separations for planets to be engulfed during the upper AGB phase of their parent stars.
The much larger radius of the AGB stars not only increases the chance for planetary interaction, but also implies that the envelope is more vulnerable to the influence of the planet towards higher final mass loss rate.
A higher mass loss rate implies also a faster post-AGB evolution. Both effects make the formation of a PN much more likely.  
our assumed low mass loss rate, therefore, supports the notion that a fraction of elliptical PNe were shaped by sub-stellar companions. 

\emph{The planetary nebula luminosity function (PNLF).}
Observations show that the brightest PNe in [O~III]~$\lambda$~5007, i.e., the bright-end cutoff of the PNLF, does not depend on the age or metallicity of the stellar population (e.g., \citealt{Ciardullo2010}).
Namely, old stellar populations for which stars of initial mass of $M_{\rm 1,ZAMS} \simeq 1-1.2 M_\odot$ are forming PNe, have their brightest [O~III]~$\lambda$~5007 PNe as young stellar populations have.
This is still a puzzle as the post-AGB stellar luminosity to ionize the nebula should be at least $\approx 5000 L_\odot$. 
Our results hint to a possible solution to this puzzle. 
As evident from the last column of Table \ref{tab:results}, in the case of a reduced mass loss rate the post-AGB stellar luminosity of stars with initial masses of $M_{\rm 1,ZAMS} \ga 1.05 M_\odot$ is $L_{\rm pAGB} \ga 5000 L_\odot$.
This, together with the interaction with a low mass companion on the upper AGB will have the necessary ingredient for a bright PN, a dense nebula of mass $\ga 0.2 M_\odot$ and a bright central ionizing star. 
  
The full solution to the puzzle might include another component. \cite{Bertolami2016} finds in a new set of simulations that the post-AGB luminosity values are higher than the values obtained in older simulations. He does not fully address the topic of the PNLF in old stellar populations,   whereas we here present preliminary results on the matter.
\cite{Mendez2017} does show a fit to the PNLF of NGC 4697 based on the \cite{Bertolami2016} tracks, yet notes that it is not a full solution to the PNLF puzzle.
It is conceivable that the new simulations of \cite{Bertolami2016} together with our assumed low mass loss rate with a late engulfment of a planet (or a brown dwarf or a low mass star), account for the brightest PNe in old stellar populations.
We point out another difference in the two works, where in contrast to \cite{Bertolami2016} we focus on binary-shaped PNe (non spherical).

\emph{The initial-final mass relation.}
There is an observed relation between the initial mass of stars and the mass of their descendant WDs 
(e.g., \citealt{Kaliraietal2008}).
Our assumed lower mass loss rate does not change much this relation for the following reasons.  
(1) Because most stars are expected to interact with stellar or sub-stellar companions, the fraction of stars that suffer no binary interaction at all is small.
(2) Non-interacting stars with low mass loss rates are supposed to account only for the more massive WD masses for each initial mass. The middle part of the final mass distribution for each initial mass, where most stars belong to,  does not change. 

\emph{The fate of the Earth.} 
Because of sensitivity to unknown tidal interaction parameters, and even to external planets \citep{Veras2016}, studies have reached contradicting conclusions about the fate of the Earth, i.e., whether
the Earth will survive engulfment (e.g., \citealt{RybickiDenis2001}), or whether the Sun will engulf the Earth (possibly already during the RGB peak of the sun; \citealt{SchroderConnonSmith2008}). 
Our assumption of a much lower mass loss rate on the giant branches, if holds for the sun, 
implies that the sun will swallow the Earth. 

\appendix
\section*{APPENDIX: Evolution with different reduced mass loss rates}
\label{app:appendix}
\setcounter{table}{0}
\renewcommand{\thetable}{A\arabic{table}}
\setcounter{figure}{0}
\renewcommand{\thefigure}{A\arabic{figure}}

\begin{figure}
\centering
\includegraphics
[trim= 3cm 6cm 4.6cm 5cm,clip=true,width=0.47\textwidth]{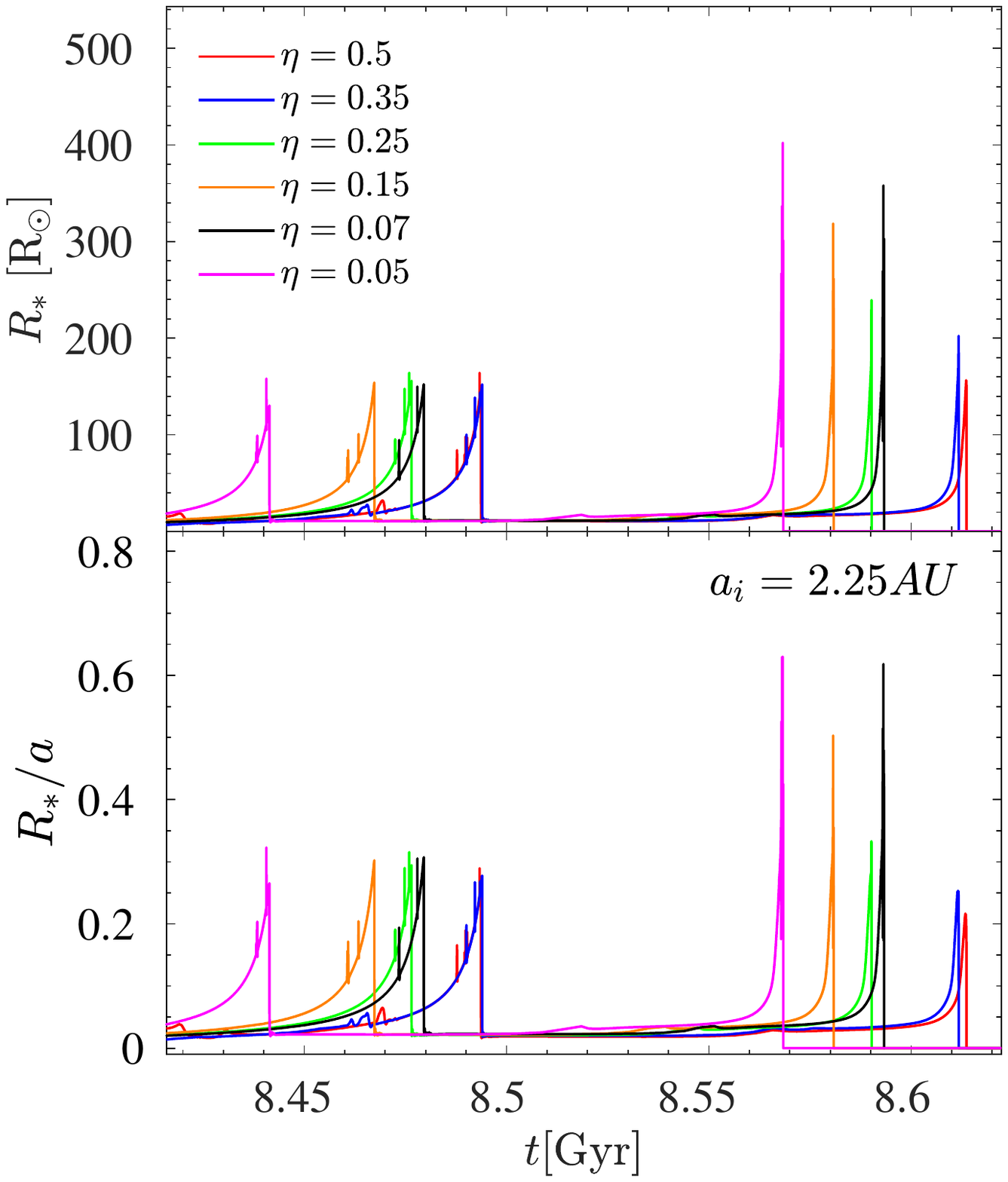} 
\hspace*{0.5cm}
\caption
{
The evolution of the radius (upper panel) and the radius over the orbital separation (lower panel) for the case of HD 159868 b, where a star of mass $1.09M_\odot$ has a planet companion of mass $2.1M_J$ at an initial semi-major axis of $a_i=2.25AU$, computed with \texttt{MESA}.
The different lines are for different values of the mass loss rate efficiency parameter $\eta$: 0.5 (red), 0.35 (blue), 0.25 (green), 0.15 (orange), 0.07 (black), and 0.05 (pink).
The upper and lower panels are as in Fig. \ref{fig:planets}.
}
\label{fig:appendix}
\end{figure}

We calculate the stellar evolution of low mass stars at an initial mass range of $M_{\rm 1,ZMAS}:0.9- 1.3M_\odot$ from ZAMS until the formation of a WD with different mass loss rate efficiency parameters: $\eta=0.5$, $0.35$, $0.25$, $0.15$, $0.07$, and $0.05$.
We study the effects of the different mass loss rates on the same four exoplanetary systems that are used in the main text (section \ref{sec:exoplanets}) and follow the same routine as in section \ref{sec:evolution}.
In Fig. \ref{fig:appendix} we present the evolution of the radius and the radius over the orbital separation for the exoplanetary system HD~159868 b, that has a $1.09M_\odot$ star, a $2.1M_J$ mass planet, and an initial semi-major axis of $a_i=2.25AU$.
It can be seen that the maximum radius on the AGB is larger than that on the RGB even for a value of $\eta$ as high as $\eta =0.15$.

In Table \ref{tab:appendix} we show the results for the 6 different mass loss rates.
As seen in our results for an efficiency parameter of $\eta\lesssim0.15$, 
the maximum value of the ratio of the stellar radius to the orbital separation (semi-major axis)  of the planet is larger when the star is on the AGB compared to that on the RGB (columns 6,7).
Moreover, the capture condition, $R_{\rm \ast, max}/a \ga 0.3$, holds on the AGB rather than on the RGB, though this is marginal for the case of the $M_{\rm 1,ZAMS}=0.9M_\odot$ star in HD 293027 A b.
In addition, the $f_{\rm AR}$ ratio (eq. \ref{eq:FAR}) is larger than unity (column 8).
These results imply that in each case that the planet will survive the RGB and will be engulfed on the AGB for a range of initial semi-major axis and the system will form an elliptical PN.

Overall a mass loss rate efficiency parameter of $\eta \la 0.15$ will be sufficient in accounting for our claims in explaining the processes of planets interacting with AGB stars and accounting for shaping non-spherical PNe.
In the main text of the paper we take a representative mass loss rate parameter of $\eta=0.07$, rather than $\eta > 0.1$, since for this value the effects we study are more pronounced and can be better explained and better compared with the effects of the evolution with the commonly used mass loss rate parameter of $\eta=0.5$. 

When examining the post-AGB luminosities (column 9 in Table \ref{tab:appendix}) we find that in order to explain the high luminosities in old stellar populations, $L_{\rm pAGB} \ga 5000 L_\odot$, our results are marginal for the lower end of our mass range.
Furthermore, an efficiency parameter of $\eta =0.15$ cannot fully produce the high luminosities needed for old stellar populations and the PNLF.
We raise the possibility that our work on reduced mass loss rates combined with the work of \cite{Bertolami2016} on stellar evolution might hold the answer to the PNLF puzzle by taking an efficiency parameter of $\eta \la 0.1$.
\begin{table*}
    \centering          
    \begin{tabular}{lccccccc}
	\multicolumn{8}{ c }{} \\	    
    \hline              
    \hline              
    Planet name & $M_i$    & $a_i$ & $m_p$  & $\eta$  & $\left(^{R_*}/_{a}\right)_{\rm max}$ &  $f_{\rm AR}$ & $L_{\rm pAGB}$
\\             &$[M_\odot]$&$[AU]$ & $[M_J]$&         &  RGB ~~ AGB     					 &               & $[L_\odot]$
     \\[1ex]
    \hline              
  HD 290327 A b &  0.9    & 3.43   & 2.5    &	0.5   &  0.15 ~~ 0.09						 & 0.65			 & $1.47\times10^3$
\\HD 290327 A b &  0.9    & 3.43   & 2.5    &	0.35  &  0.17 ~~ 0.12						 & 0.80			 & $1.87\times10^3$ 
\\HD 290327 A b &  0.9    & 3.43   & 2.5    &	0.25  &  0.17 ~~ 0.15						 & 1.02			 & $2.98\times10^3$ 
\\HD 290327 A b &  0.9    & 3.43   & 2.5    &	0.15  &  0.19 ~~ 0.25						 & 1.06			 & $3.40\times10^3$ 
\\HD 290327 A b &  0.9    & 3.43   & 2.5    &	0.07  &  0.20 ~~ 0.31						 & 1.22			 & $4.23\times10^3$ 
\\HD 290327 A b &  0.9    & 3.43   & 2.5    &	0.05  &  0.21 ~~ 0.34						 & 1.34			 & $5.31\times10^3$ 
\\    \hline             
\\47 Uma b      &  1.03   & 2.10   & 2.5    &   0.5   &  0.30 ~~ 0.21						 & 0.85			 & $1.97\times10^3$
\\47 Uma b      &  1.03   & 2.10   & 2.5    &   0.35  &  0.32 ~~ 0.26						 & 1.02			 & $2.69\times10^3$
\\47 Uma b      &  1.03   & 2.10   & 2.5    &   0.25  &  0.31 ~~ 0.32						 & 1.10			 & $3.62\times10^3$
\\47 Uma b      &  1.03   & 2.10   & 2.5    &   0.15  &  0.31 ~~ 0.48						 & 1.26			 & $4.05\times10^3$
\\47 Uma b      &  1.03   & 2.10   & 2.5    &   0.07  &  0.32 ~~ 0.60						 & 1.51			 & $4.62\times10^3$
\\47 Uma b      &  1.03   & 2.10   & 2.5    &   0.05  &  0.32 ~~ 0.66						 & 1.61			 & $5.01\times10^3$
\\    \hline             
\\HD 159868 b   &  1.09   & 2.25   & 2.1    &   0.5	  &  0.29 ~~ 0.22						 & 0.96			 & $2.18\times10^3$
\\HD 159868 b   &  1.09   & 2.25   & 2.1    &   0.35  &  0.28 ~~ 0.25						 & 1.05			 & $3.40\times10^3$
\\HD 159868 b   &  1.09   & 2.25   & 2.1    &   0.25  &  0.32 ~~ 0.33						 & 1.14			 & $1.66\times10^3$
\\HD 159868 b   &  1.09   & 2.25   & 2.1    &   0.15  &  0.30 ~~ 0.50						 & 1.30			 & $4.48\times10^3$
\\HD 159868 b   &  1.09   & 2.25   & 2.1    &   0.07  &  0.31 ~~ 0.62						 & 1.53			 & $5.24\times10^3$
\\HD 159868 b   &  1.09   & 2.25   & 2.1    &   0.05  &  0.32 ~~ 0.63						 & 1.86			 & $5.24\times10^3$
\\    \hline             
\\HD 13908  c   &  1.29   & 2.03   & 5.1    &	0.5	  &  0.29 ~~ 0.37						 & 1.20			 & $3.30\times10^3$
\\HD 13908  c   &  1.29   & 2.03   & 5.1    &	0.35  &  0.29 ~~ 0.47						 & 1.36			 & $3.59\times10^3$
\\HD 13908  c   &  1.29   & 2.03   & 5.1    &	0.25  &  0.31 ~~ 0.54						 & 1.42			 & $4.68\times10^3$
\\HD 13908  c   &  1.29   & 2.03   & 5.1    &	0.15  &  0.31 ~~ 0.64						 & 1.61			 & $4.91\times10^3$
\\HD 13908  c   &  1.29   & 2.03   & 5.1    &	0.07  &  0.32 ~~ 0.70						 & 1.81			 & $5.69\times10^3$
\\HD 13908  c   &  1.29   & 2.03   & 5.1    &	0.05  &  0.31 ~~ 0.76						 & 1.98			 & $5.72\times10^3$

\\ \hline
    \end{tabular}
\footnotesize
\flushleft
\caption{The results for the four exoplanetary systems we study.
For each system the upper line is for the commonly used mass loss rate and the lines below are the results with reduced mass loss rates of 0.35, 0.25, 0.15, 0.07, and 0.05.
From left to right: the planet name, initial mass of the primary star $M_i$, initial semi-major axis $a_i$, mass of planetary companion $m_p$ (observed parameters), the mass loss rate parameter $\eta$ (our input parameter), and then our results, the  maximum ratio of the stellar radius to the orbital separation on the RGB and on the AGB, the calculated $f_{\rm AR}$ ratio (eq. \ref{eq:FAR}) that depicts the relative importance of tidal forces on the AGB to those on the RGB, and the post-AGB luminosity of the stellar remnant. 
}
\label{tab:appendix}
\end{table*}

\section*{Acknowledgments}
We thank an anonymous referee for very helpful and detailed comments that improved the manuscript.
This research was supported by the Israel Science Foundation, by the E. and J. Bishop Research Fund at the Technion, and by the Prof. A. Pazy Research Foundation.



\label{lastpage}
\end{document}